\documentclass[aps,twocolumn,a4paper,pra,superscriptaddress,floatfix]{revtex4}%
\usepackage{amssymb}
\usepackage{amsmath}
\usepackage{amsfonts}
\usepackage{graphicx}
\usepackage{graphics}

\def\P{\mathrm{P}}    
\def\S{\mathrm{S}}    
\def\PS{\mathrm{PS}}  
\def\PFA{\mathrm{PFA}}
\def\TE{\mathrm{TE}}
\def\TM{\mathrm{TM}}
\def\bk{\mathbf{k}}
\def\dd{\mathrm{d}}
\def\E{\mathrm{E}}
\def\F{\mathrm{F}}
\def\G{\mathrm{G}}
\def\FG{\mathrm{F,G}}
\def\EF{\mathrm{E,F}}
\def\M{\mathrm{M}}
\def\cE{\mathcal{E}}
\def\cF{\mathcal{F}}
\def\cG{\mathcal{G}}
\def\cL{\mathcal{L}}
\def\cK{\mathcal{K}}
\def\cR{\mathcal{R}}
\def\cM{\mathcal{M}}
\def\max{\mathrm{max}}

\begin{document}

\title{Casimir interaction between plane and spherical metallic surfaces}

\author{Antoine Canaguier-Durand}
\affiliation{Laboratoire Kastler Brossel,
CNRS, ENS, Universit\'e Pierre et Marie Curie case 74,
Campus Jussieu, F-75252 Paris Cedex 05, France}
\author{Paulo A. Maia Neto}
\affiliation{Instituto de F\'{\i}sica, UFRJ,
CP 68528,   Rio de Janeiro,  RJ, 21941-972, Brazil}
\author{Ines Cavero-Pelaez}
\author{Astrid Lambrecht}
\author{Serge Reynaud}
\affiliation{Laboratoire Kastler Brossel,
CNRS, ENS, Universit\'e Pierre et Marie Curie case 74,
Campus Jussieu, F-75252 Paris Cedex 05, France}

\date{\today}

\begin{abstract}
We give an exact series expansion of the Casimir force between plane
and spherical metallic surfaces in the non trivial situation where
the sphere radius $R$, the plane-sphere distance $L$ and the plasma
wavelength $\lambda_\P$ have arbitrary relative values. We then present
numerical evaluation of this expansion for not too small values of
$L/R$. For metallic nanospheres where $R, L$ and $\lambda_\P$ have
comparable values, we interpret our results in terms of a
correlation between the effects of geometry beyond the proximity
force approximation (PFA) and of finite reflectivity due to material
properties. We also discuss the interest of our results for the
current Casimir experiments performed with spheres of large radius $R\gg L$.
\end{abstract}


\maketitle

The Casimir force is a striking macroscopic effect of quantum vacuum
fluctuations which has been seen in a number of dedicated
experiments in the last decade (see for example
\cite{OnofrioNJP06,DeccaPRD07} and references therein). One aim of
the Casimir force experiments is to investigate the presence of
hypothetical weak forces predicted by unification models through a
careful comparison of the measurements with quantum electrodynamics
predictions. This aim can only be reached if theoretical
computations are able to take into account a realistic and reliable
modeling of the experimental conditions. Among the effects to be
taken into account are the material properties and the surface
geometry, these effects being also able to produce phenomena of
interest in nanosystems \cite{Chan08,Lambrecht08}.

A number of Casimir measurements have been performed with
gold-covered plane and spherical surfaces separated by distances $L$
of the order of the plasma wavelength ($\lambda_\P\simeq136$nm for
gold), making material properties important in their analysis
\cite{Lambrecht00}. As those measurements use spheres with a radius
$R\gg L$, they are commonly analyzed through the Proximity Force
Approximation (PFA) \cite{Derjaguin68}, which amounts to a trivial
integration over the sphere-plate distances. An exception is the
Purdue experiment dedicated to the investigation of the accuracy of
PFA in the sphere-plate geometry \cite{Krause07}, the result of
which will be given as a precise statement below.

In the present letter, we give for the first time an exact series
expansion of the Casimir force between a plane and a sphere in
electromagnetic vacuum, taking into account the material properties
via the plasma model (see Fig.~1). We present numerical evaluation
of this expansion which are limited to not too small values of
$L/R$, because of the multipolar nature of the series. We show below
that these new results lead to a striking correlation between the
effects of geometry and imperfect reflection when evaluated for
nanospheres, with $R, L$ and $\lambda_\P$ having comparable values.
In the end of this letter, we also discuss the interest of these
results for the Casimir experiments performed with large spheres
$R\gg L$ \cite{Krause07}.

\begin{figure}[h]
\centering
\includegraphics[width=3cm]{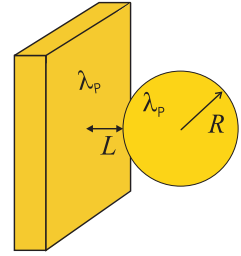}
\caption{The geometry of a sphere of radius $R$ and a flat plate at
a distance $L$ (center-to-plate distance $\cL\equiv L+R$); both
mirrors are covered with a metal characterized by a plasma
wavelength $\lambda_\P$.}
\end{figure}

Our starting point is a general scattering formula for the Casimir
energy \cite{LambrechtNJP06}. Using suitable plane-wave and multipole
bases, we deduce the Casimir energy $\cE_\PS$ between a plane and a
spherical metallic surface in electromagnetic vacuum. The multipole
series expansion is written in terms of Fresnel reflection
amplitudes for the plate and Mie coefficients for the sphere, and it
is valid for arbitrary relative values of the sphere radius $R$, the
sphere-plate distance $L$ and the plasma wavelength $\lambda_\P$.
For the sake of comparison with experiments, we assume
$\lambda_\P\simeq136$nm for both, the sphere and the plate. We 
occasionally also consider the limit $\lambda_\P\rightarrow0$, where
the formula reduces to the case of perfect reflectors in
electromagnetic vacuum, for which results were obtained recently
\cite{Emig08,Neto08,Kenneth08}.

In the following, we discuss the force
$\cF_\PS\equiv-\partial\cE_\PS/\partial L$ as well as the force
gradient $\cG_\PS\equiv-\partial\cF_\PS/\partial L$ which was
measured in the experiment \cite{Krause07}. We write the results
deduced from the scattering formula as products of PFA estimates by
beyond-PFA correction factors $\rho_\F$ and $\rho_\G$:
\begin{eqnarray}
\label{defrho}
\cF_\PS \equiv \rho_\F \cF_\PS^\PFA &\quad,\quad&
\cF_\PS^\PFA\equiv\eta_\E \frac{\hbar c\pi^3 R}{360L^3}\nonumber\\
\cG_\PS \equiv \rho_\G \cG_\PS^\PFA &\quad,\quad&
\cG_\PS^\PFA\equiv\eta_\F \frac{\hbar c\pi^3 R}{120L^4}
\end{eqnarray}
The PFA estimates $\cF_\PS^\PFA$ and $\cG_\PS^\PFA$ are
proportional respectively to the energy and force calculated between
two planes. They are written as products of ideal Casimir
expressions and factors $\eta_\E$ and $\eta_\F$ accounting for the
effect of imperfect reflection \cite{Lambrecht00}.

The beyond-PFA correction factors $\rho_\F$ and $\rho_\G$ appearing
in (\ref{defrho}) are the important quantities for what follows. For
experiments performed with large spheres of radius $R\gg L$, the deviation
from PFA is small ($\rho_\F\simeq1$). Even in this limit, it remains
important to specify the accuracy of PFA in order to master the
quality of theory-experiment comparison \cite{Neto08}. This can be
done by introducing a Taylor expansion of the correction factors at
small values of $L/R$
\begin{eqnarray}
\label{develrho}
&&\rho_\FG \equiv 1 + \beta_\FG \ \frac{L}{R} + O\left( \frac{L^2}{R^2} \right)
\end{eqnarray}
The only experimental result available on this topic \cite{Krause07}
may be stated as a bound on the $\beta_G$ factor, namely
$\vert\beta_\G\vert<0.4$. On the theoretical side, analytical as
well as numerical calculations of this slope have been obtained for
scalar field models
\cite{Schaden00,Gies06,Jaffe06,Wirzba08,Bordag08}. 
For the situation met in experiments, with a plane and a sphere in
electromagnetic vacuum, an estimation technique has recently been proposed
where the slope is deduced from a polynomial fit of the numerical values
obtained at intermediate values of $L/R$ \cite{Emig08,Neto08}.
 The slope obtained in this
manner is much larger ($\sim$8 times larger) than expected from scalar field
models \cite{Neto08}.
As a consequence, the value of
$\beta_G$ falls out of the bound of \cite{Krause07}, in contrast
with the scalar prediction which lies within the bound. More precise
statements on this point will be given below.

On the other hand all these results correspond to perfect
reflection, whereas the experiment \cite{Krause07} was performed
with gold-covered surfaces. The apparent contradiction noticed in the
preceding paragraph may thus be cured if the value of $\beta_G$
differs for metallic and perfect mirrors, that is also if the
effects of geometry and finite reflectivity are correlated. We show
in the sequel of the letter that this is indeed the case.

We start from the formula for the Casimir energy $\cE_\PS$ between
two scatterers in vacuum \cite{LambrechtNJP06}
\begin{eqnarray}
\label{depart}
&&\cE_\PS=\hbar\,\int_0^\infty \frac{\dd\xi}{2\pi}\log{\rm det}
\left(1- \cM \right) \nonumber \\
&&\cM\equiv \cR_\S e^{-\cK \cL} \cR_\P e^{-\cK \cL}
\end{eqnarray}
In the geometry depicted on Fig.~1 with a sphere of radius $R$, a
plate, and a sphere-plate separation $L$ along the $z$-axis
(center-to-plate distance $\cL\equiv L+R$), $\cR_\S$ and $\cR_\P$
represent the reflection operators for the spherical and the plane
scatterers, respectively. They are evaluated with reference points
placed at the sphere center and at its projection on the plane,
respectively. The operator $e^{-\cK \cL}$ describes the one-way
propagation between these two reference points.
$\xi$ is the imaginary field frequency integrated over the upper
imaginary axis.

In order to evaluate explicitly this expression, we use two mode
decompositions. The first one is a plane-wave basis
$\left\vert\bk,\phi,p \right\rangle_\xi$ with $\bk$ the transverse
wavevector parallel to the $xy$ plane, $p=\TE,\TM$ the polarization,
and $\phi=\pm 1$ for rightward/leftward propagation directions. It
is well adapted to the description of free propagation and
reflection on the plane: the propagation operator $e^{-\cK \cL}$ is
diagonal with matrix elements $e^{-K \cL}$ such that $K =
\sqrt{\xi^2/c^2+k^2}$ ($k\equiv\vert\bk\vert$) while reflection on
the plane preserves all plane-wave quantum numbers but $\phi$. The
non zero elements of  $\cR_\P$ are the standard Fresnel reflection
amplitudes $r_p$. Given values of $\bk(k,\varphi)$ and $\phi=\pm 1$
define a direction in reciprocal space corresponding to the
azimuthal angle $\varphi$ and a complex angle $\theta^\pm$ such that
$\sin\theta^\pm=-i\frac{ck}\xi$ and
$\cos\theta^\pm=\pm\frac{cK}\xi$.

The second basis, which is adapted to the spherical symmetry of
$\cR_\S$, is a  multipole basis $\left\vert\ell m
P\right\rangle_\xi$, with $\ell(\ell+1)$ and $m$ the angular
momentum eigenvalues ($\ell=1,2,...$, $m=-\ell,...,\ell$) and
$P=\E,\M$ for the electric and magnetic multipoles. By rotational
symmetry around the $z$-axis, $\cM$ commutes with $J_z$. Hence it is
block diagonal, with each block $\cM^{(m)}$ corresponding to a
common value of $m$ and yielding a contribution $\cE_\PS^{(m)}$ to
the Casimir energy $\cE_\PS$ (opposite values $\pm m$ provide
identical contributions). The contribution $\cE_\PS^{(m)}$ is
written as in (\ref{depart}) with $\cM$ replaced by the block matrix
\begin{equation}
\label{blocks}
\cM^{(m)} = \left(
\begin{array}
[c]{cc}%
M^{(m)}(\E,\E)
 & M^{(m)}(\E,\M)
\\
M^{(m)}(\M,\E) & M^{(m)}(\M,\M)
\end{array}
\right)
\end{equation}
Each block in this matrix is the sum of  TE and TM contributions
$M^{(m)}(P_1,P_2)=\sum_p M^{(m)}_{p}(P_1,P_2).$ The diagonal blocks
are written as
\begin{eqnarray}
\label{diagonal}
& M^{(m)}_{\TE}(\E,\E)_{\ell_1,\ell_2} =
   \sqrt{\frac{\pi(2\ell_1+1)}{\ell_2(\ell_2+1)}}
\,A^{(m)}_{\ell_1,\ell_2,\TE}\,a_{\ell_1}(i\xi)
\nonumber \\
& M^{(m)}_{\TM}(\E,\E)_{\ell_1,\ell_2} =
    \sqrt{\frac{\pi(2\ell_1+1)}{\ell_2(\ell_2+1)}}
\,B^{(m)}_{\ell_1,\ell_2,\TM}\,a_{\ell_1}(i\xi)
\nonumber \\
&M^{(m)}_{\TM}(\M,\M)_{\ell_1,\ell_2} =
    \sqrt{\frac{\pi(2\ell_1+1)}{\ell_2(\ell_2+1)}}
\,A^{(m)}_{\ell_1,\ell_2,\TM}\,b_{\ell_1}(i\xi)
\nonumber \\
& M^{(m)}_{\TE}(\M,\M)_{\ell_1,\ell_2} =
  \sqrt{\frac{\pi(2\ell_1+1)}{\ell_2(\ell_2+1)}}
\,B^{(m)}_{\ell_1,\ell_2,\TE}\,b_{\ell_1}(i\xi)
\end{eqnarray}
$a_{\ell}(i\xi)$ and $b_{\ell}(i\xi)$ are the Mie coefficients \cite{Bohren}
for electric and magnetic multipoles.
$A$ and $B$ are matrices which do not depend on the radius nor on the refractive
index of the sphere and are written in terms of the spherical harmonics
$Y_{\ell,m}(\theta,\varphi=0)$ and the finite rotation matrix elements
$d^{\ell}_{m,m'}(\theta)= \langle \ell, m| e^{-i\theta J_y} |\ell, m'\rangle$
\cite{Edmonds}
\begin{eqnarray}
\label{AB}
 A^{(m)}_{\ell_1,\ell_2,p} = -i m
\int_0^\infty \frac{\dd k}{K}\left(d^{\ell_1}_{m,1}\left(\theta^+\right)+
d^{\ell_1}_{m,-1}\left(\theta^+\right)\right)
\nonumber \\
\times Y_{\ell_2m}\left(\theta^-\right)\, r_p(k)\, e^{-2K \cL}
\nonumber \\
 B^{(m)}_{\ell_1,\ell_2,p} =  -\frac{c}{\xi}
\int_0^\infty \frac{k\dd k}{K}\left(d^{\ell_1}_{m,1}\left(\theta^+\right)-
d^{\ell_1}_{m,-1}\left(\theta^+\right)\right)
\nonumber \\
\times \partial_{\theta} Y_{\ell_2m}\left(\theta^-\right)\, r_p(k)\, e^{-2K \cL}
\end{eqnarray}
Similar expressions are found for the nondiagonal blocks, with
the matrices $A$ and $B$ replaced respectively by
\begin{eqnarray}
\label{CD}
 C^{(m)}_{\ell_1,\ell_2,p} =   \frac{c}{\xi}
\int_0^\infty \frac{k \dd k}{K}\left(d^{\ell_1}_{m,1}\left(\theta^{+}\right)+
d^{\ell_1}_{m,-1}\left(\theta^{+}\right)\right)
\nonumber\\
\times \partial_{\theta} Y_{\ell_2m}\left(\theta^{-}\right)\, r_p(k)\, e^{-2K \cL}
\nonumber\\
 D^{(m)}_{\ell_1,\ell_2,p} =  i  m
\int_0^{\infty} \frac{\dd k}{K}\left(d^{\ell_1}_{m,1}\left(\theta^{+}\right)-
d^{\ell_1}_{m,-1}\left(\theta^{+}\right)\right)
\nonumber\\
\times  Y_{\ell_2m}\left(\theta^{-}\right)\, r_p(k)\, e^{-2K \cL}
\end{eqnarray}

In order to go further, we assume the materials to have a dielectric
response described by the plasma model $\epsilon(i\xi)=
1+\omega_\P^2/\xi^2$, with $\omega_\P$ the plasma frequency and
$\lambda_\P=2\pi c/\omega_\P$ the plasma wavelength. Although the
formalism easily allows for different values of $\lambda_\P$ for
both surfaces, we take a common value as in the recent
experiment \cite{Krause07}. We calculate the Casimir energy $\cE_\PS$
and deduce the force $\cF_\PS$ and gradient $\cG_\PS$, both
quantities being functions of the 3 length scales $R$, $L$ and
$\lambda_\P$. The case of perfect reflection \cite{Neto08} can be
recovered as the limit $\lambda_\P\ll R,L$ (see \cite{Noguez} for
the opposite non retarded limit). A large distance limit may also be
taken as $\lambda_\P,R \ll L$. Its result reduces to the Rayleigh
expression \cite{Buhman} in the case ($R \ll \lambda_\P$) or to
$3/2$ of it \cite{Emig08,Neto08} in the case ($\lambda_\P\ll R$).

\begin{figure}[b]
\centering
\includegraphics[width=7cm]{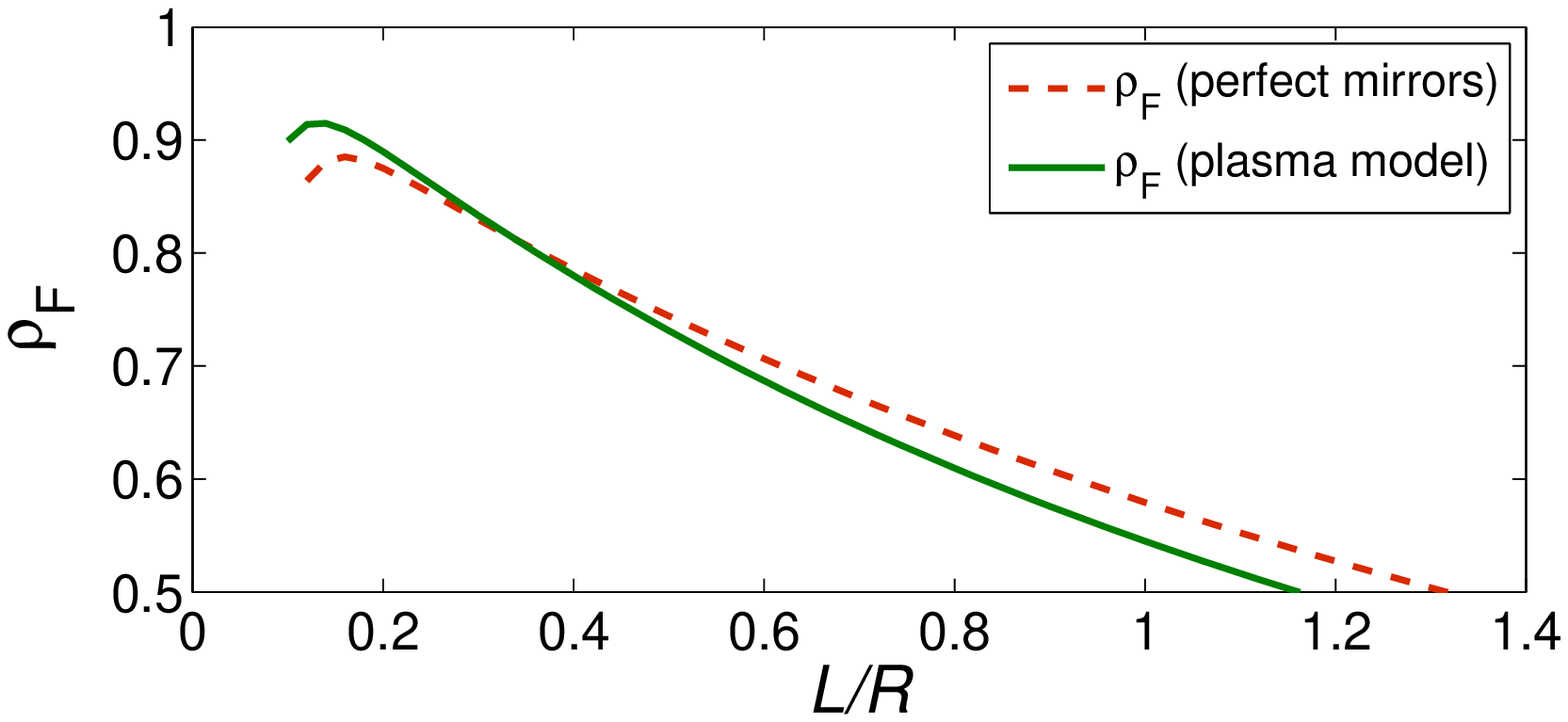}
\includegraphics[width=7cm]{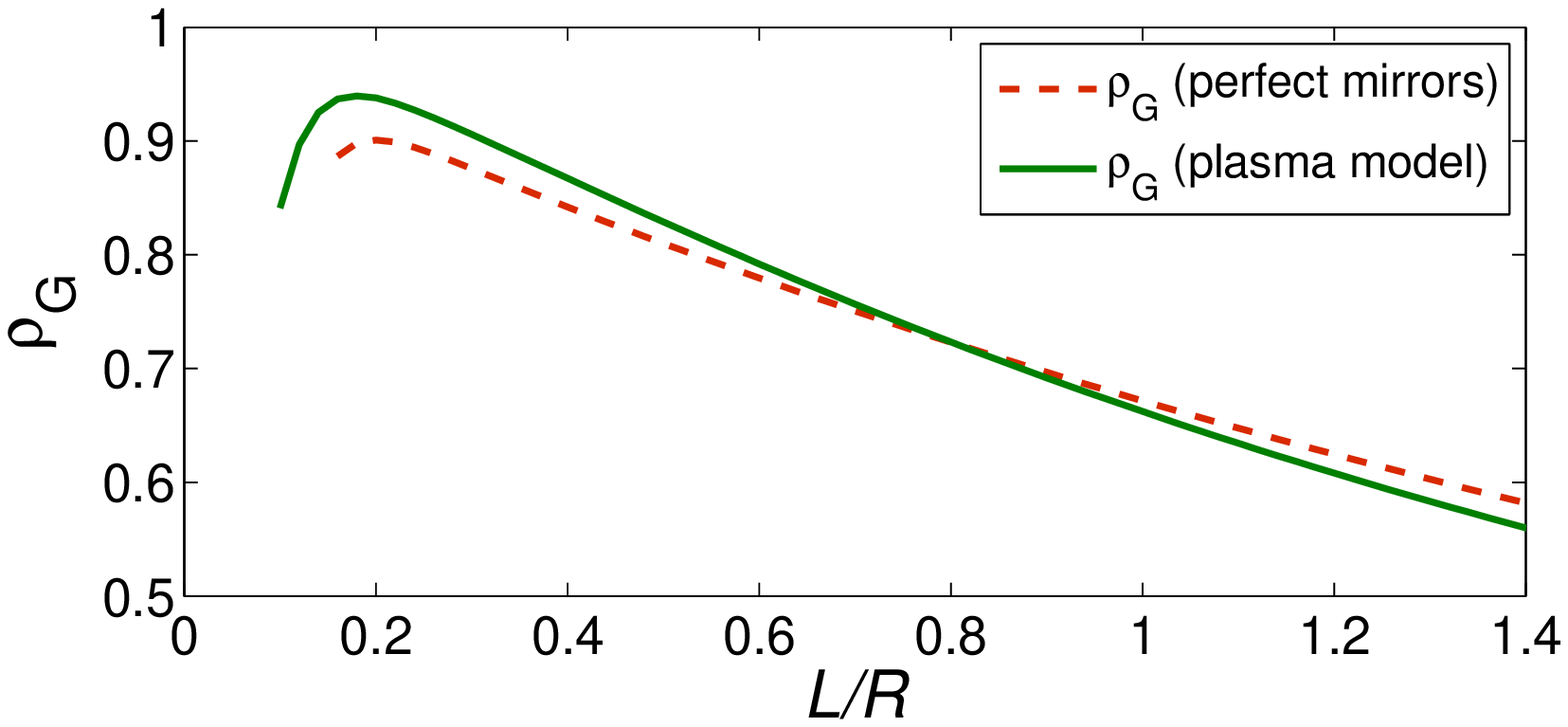}
\caption{Upper graph : variation of $\rho_\F$ as a function of
$L/R$, for a nanosphere of radius $R=100$nm; the solid green line
corresponds to gold-covered plates ($\lambda_\P=136$nm) and the
dashed red line to perfect reflectors. Lower graph : variation of
$\rho_\G$ as a function of $L/R$, with the same conventions as on
upper graph. The decreases at low values of $L/R$ represent a
numerical inaccuracy due to the limited value of $\ell_\max=24$
[Colors online].} \label{rhoFG_perfect_plasma}
\end{figure}

As already discussed, the PFA expression is also contained in our
general result, and it is recovered asymptotically for $R \gg L$. In
the following, we discuss the results of numerical computations of
the ratios $\rho_\FG$ defined in (\ref{defrho}) which measure the
deviation from PFA. For dimensionality reasons $\rho_\FG$ are
functions of two dimensionless parameters built upon $L, R$ and
$\lambda_\P$ ($\eta_\EF$ are functions of $L/\lambda_\P$ only
\cite{Lambrecht00}) and they approach unity at the PFA limit
$L/R\ll1$. Their numerical computation is done after truncating the
vector space at some maximum value $\ell_\max$ of the orbital number
$\ell$. As a consequence of the `localization principle'
\cite{Moyses}, the results are accurate only for $R/L$ smaller than
some value which increases with $\ell_\max$. At the moment, our
numerical calculations are limited to $\ell_\max=24$, allowing us to
obtain accurate results down to $ L/R~\simeq 0.2$ but not in close
vicinity of the PFA limit.

This method gives new and interesting results, in particular for
nanospheres having a radius $R$ with the same order of magnitude as
the plasma wavelength $\lambda_\P.$ In this case, we can perform accurate
calculations for $L$ having a comparable magnitude, and thus explore
the rich functional dependence of $\rho_\FG$ versus two
dimensionless parameters built up on $L, R$ and $\lambda_\P$.
Fig.~\ref{rhoFG_perfect_plasma} shows the results obtained for
$\rho_\F$ and $\rho_\G$ with metallic and perfect mirrors. Clearly
the deviation from PFA calculated for metallic mirrors differs
markedly from that already known for perfect mirrors. For small
values of $L/R$ the violation of PFA for the Casimir force and
gradient turns out to be less pronounced for metallic mirrors than
for perfect mirrors, while for large values of $L/R$ it is more
pronounced.

\begin{figure}[b]
\centering
\includegraphics[width=7cm]{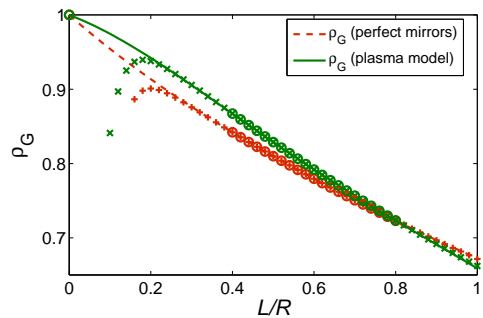}
\caption{Quartic polynomial fit of the function $\rho_\G(L/R)$,
for a nanosphere of radius $R=100$nm;
the solid green line corresponds to gold-covered plates
and the dashed red line to perfect reflectors.
The crosses represent numerically evaluated points
and the circles indicate those points which are
used for the fit [Colors online].}
\label{rhoG_perfect_plasma_fit}
\end{figure}

However, at values $L/R \simeq 0.2$ we find a clear correlation
between geometry and finite reflectivity effects, making
therefore measurements with nanospheres at small plate-sphere
separations particularly interesting. This non trivial interplay
becomes evident when a polynomial fit of the numerical values of
$\rho_\FG$ is used for inferring the behaviour at small values of
$L/R$ \cite{Emig08,Neto08}. On Fig.~\ref{rhoG_perfect_plasma_fit} we
plot the quartic polynomial fits of the function $\rho_G$ for the
two cases of gold-covered and perfect mirrors. The curves were
obtained by finding the best-fit of the numerically computed values
of $\rho_G$ (crosses on Fig.~\ref{rhoG_perfect_plasma_fit}) in the
window $0.4<L/R<0.8$ (circled crosses on
Fig.~\ref{rhoG_perfect_plasma_fit}) in the set of quartic
polynomials (Taylor expansion defined as in (\ref{develrho}) and
truncated at fourth order). The lefthand bound of the window is
fixed by the requirement of using only points accurately calculated
with $\ell_\max = 24$ while the righthand bound is determined by the
truncation at fourth order of the Taylor expansion. The best-fits
correspond to the following polynomials for gold-covered (GM) and
perfect (PM) mirrors respectively ($x\equiv L/R$)
\begin{eqnarray}
\label{bestfit100nano}
\mathrm{GM}&:& 1 - 0.207 x  - 0.530 x^2 + 0.645 x^3 - 0.249 x^4 \nonumber\\
\mathrm{PM}&:& 1 - 0.483 x  + 0.297 x^2 - 0.221 x^3 + 0.080 x^4
\end{eqnarray}
The two fits are clearly different and this in particular the case
for the values obtained for the slope at $L/R=0$. The slope
($\beta_\G\sim-0.21$) obtained for gold mirrors differs by more than
a factor 2 from the one ($\beta_\G\sim-0.48$) obtained for perfect
mirrors. This is related to the bending of the curve for gold
mirrors at small $L/R$, which describes the effect of imperfect
reflection in the beyond-PFA factor $\rho_\G$ and has to be
contrasted with the unbent curve for perfect mirrors. For the same
reason, we observe that the slope obtained for gold mirrors is less
stable under the variation of the conditions of the best-fit
procedure than that for perfect mirrors. To appreciate the meaning
of the bending let us recall that the slope obtained for perfect
mirrors in an electromagnetic vacuum is $\sim$8 times larger than
expected from scalar computations \cite{Wirzba08,Bordag08} and one
cannot but notice that it lies outside the bound
$\vert\beta_\G\vert<0.4$ of \cite{Krause07}. In contrast, the slope
obtained for metallic mirrors lies within the bound. Let us
emphasize that there is no contradiction between the results
presented here (obtained for nanospheres with $R=100$nm) and the
experiments (performed with spheres having $R>$ a few tenths of
$\mu$m).

For  spheres with  large radii ($L/R>0.2$) the beyond-PFA factors
$\rho_\FG$ have the same values for gold-covered and perfect
mirrors, because the value of $L$ is much larger than $\lambda_\P$.
If we extracted a slope from these results, we would obtain a value
close to that of perfect mirrors, thus lying outside the bound of
\cite{Krause07}. However, the arguments discussed before show that
one should refrain from doing so. Indeed, a bending of the curve has
to be expected in this case too, for values of $L$ becoming
comparable to $\lambda_\P$ and thus much smaller than $R$. In
contrast, this bending has no reason to appear for perfect mirrors
since there is no length scale like $\lambda_\P$ in this case. If
the bending is similar for large and small spheres, it may turn out
that the slope for gold-covered mirrors meets the bound
\cite{Krause07} while that for perfect mirrors does not.

To sum up our results, we have written a new and exact expansion for
the Casimir force between plane and spherical metallic surfaces in
electromagnetic vacuum. The results go beyond the proximity force
approximation, and show a clear correlation between the plane-sphere
geometry and the material properties of the metallic surfaces. They
constitute a new step in the direction of accurate comparisons
between Casimir experiments and QED theoretical predictions. More
work is needed to obtain exact results for the Casimir force between
a metallic sphere and plate in the so far experimentally explored
parameter region of $L/R \simeq 0.01$, using for example different
approaches based on semiclassical methods. Our results also
indicate a complementary way to observe deviations from PFA
and the interplay between geometrical and reflectivity effects in
new experiments performed with nanospheres.

\acknowledgements The authors thank M.T. Jaekel, C. Genet, D.A.R.
Dalvit, D. Delande, B. Gremaud and V. Nesvizhevsky for stimulating
discussions. P.A.M.N. thanks  CNPq, CAPES, Institutos do Mil\^enio
de Informa\c c\~ao Qu\^antica e Nanoci\^encias for financial support
and ENS for a visiting professor position. I.C.P and A.L.
acknowledge financial support from the French Carnot Institute LETI.

\newcommand{\REVIEW}[4]{\textrm{#1} \textbf{#2}, #3, (#4)}
\newcommand{\Review}[1]{\textrm{#1}}
\newcommand{\Volume}[1]{\textbf{#1}}
\newcommand{\Book}[1]{\textit{#1}}
\newcommand{\Eprint}[1]{\textsf{#1}}
\def\etal{\textit{et al}}

\end{document}